\documentclass{aa}
\usepackage{graphicx}

\newcommand{\kmprs}  {\mbox{\rm km\,s$^{-1}$}}

\newcommand{\Vtot}  {\mbox{$V_{\rm tot}$}}

\newcommand{\feh} {\mbox{\rm [Fe/H]}}
\newcommand{\nih} {\mbox{\rm [Ni/H]}}
\newcommand{\znh} {\mbox{\rm [Zn/H]}}
\newcommand{\ZnFe} {\mbox{\rm [Zn/Fe]}}
\newcommand{\SFe} {\mbox{\rm [S/Fe]}}
\newcommand{\SZn} {\mbox{\rm [S/Zn]}}
\newcommand{\ZnNi} {\mbox{\rm [Zn/Ni]}}
\newcommand{\NiFe} {\mbox{\rm [Ni/Fe]}}
\newcommand{\NaFe} {\mbox{\rm [Na/Fe]}}
\newcommand{\OFe} {\mbox{\rm [O/Fe]}}
\newcommand{\MgFe} {\mbox{\rm [Mg/Fe]}}
\newcommand{\CaFe} {\mbox{\rm [Ca/Fe]}}
\newcommand{\teff}  {\mbox{$T_{\rm eff}$}}

\newcommand{\logg}  {\mbox{{\rm log}\,$g$}}
\newcommand{\FeI} {\ion{Fe}{i}}
\newcommand{\FeII} {\ion{Fe}{ii}}
\newcommand{\ZnI} {\ion{Zn}{i}}
\newcommand{\ZnII} {\ion{Zn}{ii}}
\newcommand{\CaI} {\ion{Ca}{i}}
\newcommand{\NiI} {\ion{Ni}{i}}
\newcommand{\ffe}  {\mbox{${\rm [\frac{Fe}{H}]}$}}
\newcommand{\znfe}  {\mbox{${\rm [\frac{Zn}{Fe}]}$}}
\newcommand{\nife} {\mbox{${\rm [\frac{Ni}{Fe}]}$}}

\begin{document}

\title{The [Zn/Fe] -- [Fe/H] trend for disk and halo stars
\thanks{Based on observations collected at the
National Astronomical Observatories, Xinglong, China
and the European Southern Observatory, La Silla, Chile
(ESO No. 67.D-0106).}}

\author{Y.Q.~Chen \inst{1} \and P.E.~Nissen \inst{2}
\and G.~Zhao \inst{1}}

\offprints{Y.Q.~Chen}

\institute{
National Astronomical Observatories, Chinese Academy of Sciences,
   Beijing 100012, P.R. China
\email{cyq@bao.ac.cn,zg@bao.ac.cn}
\and
Institute of Physics and Astronomy, University of Aarhus, DK--8000
Aarhus C, Denmark.
\email{pen@phys.au.dk} }

\date{Received ..... / Accepted ......}

\abstract{Zn abundances, derived from a model atmosphere
analysis  of the $\lambda$6362.35\,\AA\
\ZnI\ line, are presented for 44 thin disk, 10 thick disk and  8
halo dwarf stars in the metallicity range $-1.0 < \feh < +0.2$.
It is found that \ZnFe\ in thin disk
stars shows a slight increasing trend with decreasing metallicity
reaching a value $\ZnFe \simeq \! +0.1$ at $\feh = -0.6$. 
The thick disk stars in the metallicity range $-0.9 < \feh < -0.6$ 
have an average $\ZnFe\ \simeq \! +0.15$\,dex, whereas five 
alpha-poor and Ni-poor halo stars in the same metallicity range 
have $\ZnFe\ \simeq \! 0.0$\,dex. 
These results indicate that Zn is not an exact tracer of Fe as often
assumed in abundance studies of damped Lyman-alpha systems (DLAs).
A better understanding of the nucleosynthesis of Zn is needed
in order to obtain more detailed information on the past history
of star formation in DLAs from e.g. the observed sulphur/zinc ratio.
} 

\maketitle

\section{Introduction}
\label{intro}
The determination of zinc abundances in Galactic stars is of high
interest in astrophysics for at least two reasons. Firstly, 
the nucleosynthesis of zinc is not well understood.
According to Woosley \& Weaver (\cite{Woosley95})
Zn is produced by supernovae (SNe) of
Type II from two processes: i) neutron capture on iron group
nuclei during He and C-burning 
(the weak $s$-process; Langer et al. \cite{Langer89}),
and ii) alpha-rich freeze-out following nuclear statistical 
equilibrium in layers heated to more than 5 $\times 10^9$\,K. 
However,
as shown by Timmes et al. (\cite{Timmes95}), and Goswami \& Prantzos
(\cite{Goswami00}), the corresponding yields underpredicts Zn/Fe
by about a factor of two compared to the observed ratio in Galactic
halo stars. Furthermore, the yield corresponding to the first
process is metallicity dependent leading to a predicted rise of \ZnFe\ 
as a function of [Fe/H] for disk stars in disagreement with
the rather flat trend of  \ZnFe\ vs. \feh\ observed by 
Sneden et al. (\cite{Sneden91}). In order to get a better agreement 
with the observed trend, Matteucci et al. (\cite{Matteucci93})
argued that Type Ia SNe give a very significant contribution
to the production of Zn, but according to Iwamoto et al. (\cite{Iwamoto99})
standard models of Type Ia SNe produce little Zn. As an alternative source
of Zn, Hoffman et al. (\cite{Hoffman96}) suggested that a significant 
amount of zinc (together with light $p$-process nuclei) could be
produced in the neutrino-powered wind of 10 to 20 solar mass SNe,
essentially the same site as proposed for the $r$-process
(Woosley et al. \cite{Woosley94}) but at earlier times after the explosion.
Furthermore, Umeda \& Nomoto (\cite{Umeda02}) have discussed how
the produced Zn/Fe ratio in massive metal-poor SNe depends
on the mass cut, neutron excess and explosion energy, in an attempt
to explain the high \ZnFe\ values observed for the most metal-poor stars
(Primas et al. \cite{Primas00}, Cayrel et al. \cite{Cayrel04}). 

Secondly, zinc is a key element in studies of abundances in
damped Ly$\alpha$ (DLA) systems, because it is the only element 
in the iron-peak group, which is undepleted onto dust in the
interstellar medium. With reference to the data of 
Sneden et al. (\cite{Sneden91}),
Zn is often taken as a proxy for Fe in DLA studies to derive
dust depletion factors (e.g. Vladilo \cite{Vladilo02}) and to date the 
the star formation process at high $z$ from [$\alpha$/Fe], i.e. the
logarithmic ratio between the abundance of alpha-elements 
(O, Ne, Mg, Si, S, Ca, Ti) and the abundance of iron
(e.g. Pettini et al. \cite{Pettini99},
Centuri\'{o}n et al. \cite{Centurion00}). Given the
uncertainty about the nucleosynthetic origin of zinc doubts have, 
however, been raised about the reliability of this method
(Prochaska \cite{Prochaska03}, Fenner et al. \cite{Fenner04}).
 
The often cited conclusion by Sneden et al. (\cite{Sneden91}) that
zinc abundances closely track the overall metallicity with no 
discernible change in [Zn/Fe] in the range $-2.9 < \feh  < -0.1$ has been
challenged in a number of recent studies. Prochaska et al. (\cite{Prochaska00})
found a mild enhancement of Zn relative to Fe ($\ZnFe \sim +0.1$) in ten 
thick disk stars with metallicities between $-1.2$ and $-0.4$. 
Mishenina et al. (2002) have published a survey of Zn abundances 
in 90 disk and halo stars, and although they conclude that the data 
``confirms the well-known fact that \ZnFe\ is almost solar at all 
metallicities'', Nissen (\cite{Nissen04b}) notes that there is
a tendency for thick disk stars in the metallicity range $-1.0 < \feh < -0.5$
to be overabundant in Zn. In addition,
Reddy et al. (\cite{Reddy03}) have found a slightly increasing \ZnFe\ with
decreasing metallicity for 181 thin disk stars in the metallicity range 
$-0.8 < \feh < +0.1$, and Bensby et al. (\cite{Bensby03}) found
evidence of a separation of \ZnFe\ between thin and thick disk stars as
well as a tendency of an uprising \ZnFe\ at $\feh > 0$.
Among halo stars, Nissen et al. (\cite{Nissen04b}) derived $\ZnFe \sim +0.1$
for stars in the metallicity range $-2.5 < \feh < -2.0$, and as noted above
Primas et al. (\cite{Primas00}) and Cayrel et al. (\cite{Cayrel04}) have
found clear evidence of increasing \ZnFe\ values below $\feh \simeq -2.5$
with \ZnFe\ reaching a value of +0.5 dex at $\feh \simeq -4.0$.

In most of these works, the zinc abundances are primarily
based on the $\lambda \lambda$\,4722.16, 4810.54\,\AA\ \ZnI\ lines, 
although the weak \ZnI\ line at 6362.35\,\AA\ is included in some
of the disk star studies. A problem with the 
$\lambda \lambda$\,4722.16, 4810.54\,\AA\ \ZnI\ lines is that they are
rather strong at solar metallicities ($EW \sim 70 - 80$\,m\AA\ in the
solar flux spectrum) and blended by several weak lines in the wings.
This makes it difficult to set a reliable continuum and to measure their
equivalent widths accurately. The derived Zn abundances depend furthermore
critically on the assumed value of the Van der Waals damping
constant. This means that the trend of \ZnFe\ versus \feh\ for disk stars
becomes rather uncertain if the $\lambda \lambda$\,4722.16, 4810.54\,\AA\ 
\ZnI\ lines are included in the analysis.

In the present paper we derive zinc abundances exclusively from the 
weak  6362.35\,\AA\ \ZnI\ line, for which the equivalent width ranges
from a few m\AA\ in the most metal-poor disk stars to about 30\,m\AA\
in the metal-rich stars. Hence, the line is rather ideal for 
accurate abundance determinations being insensitive to microturbulence
and Van der Waals damping parameters. In the following Sect. \ref{obs}, 
high resolution observations of this line is presented. Sect.
\ref{analysis} describes the model atmosphere analysis of the data
and possible errors are discussed in Sect. \ref{errors}. The results 
are discussed in Sect. \ref{results}.

\section{Observations and data reduction}
\label{obs}
Two sets of stars are included in this paper. The first set consists
mainly of thin disk stars from Chen et al. (\cite{Chen00}, \cite{Chen02}),
who determined abundances of solar-type dwarfs selected from
the $uvby - \beta$ photometric catalogues of Olsen (\cite{Olsen83},
\cite{Olsen93}, \cite{Olsen94}). The stars have 
$5600 \leq \teff \leq 6400$~K, $\logg \geq 3.8$
and $-1.0 \leq \feh \leq +0.3$. 
In Chen et al. (\cite{Chen00}) chemical abundances of O, Na, Mg, Al,
Si, K, Ca, Ti, V, Cr, Fe, Ni and Ba were derived from high-resolution
spectra of 90 disk stars. Thirty of these spectra are of sufficiently
high quality to allow measurements of the equivalent width of the 
6362.35\,\AA\ \ZnI\ line
with good accuracy. In Chen et al. (\cite{Chen02}) S, Si and Fe abundances
were derived for 26 disk stars; 14 of these are included in the present
paper. As described in the above papers, the spectra were
obtained with the Coud\'{e} Echelle Spectrograph attached to the
2.16m telescope at the National Astronomical Observatories (Xinglong, China).
The resolution is 37\,000 and the signal-to-noise ratio
is generally over 200 in the region of the 6362.35\,\AA\ \ZnI\ line.

\begin{figure}
\resizebox{\hsize}{!}{\includegraphics{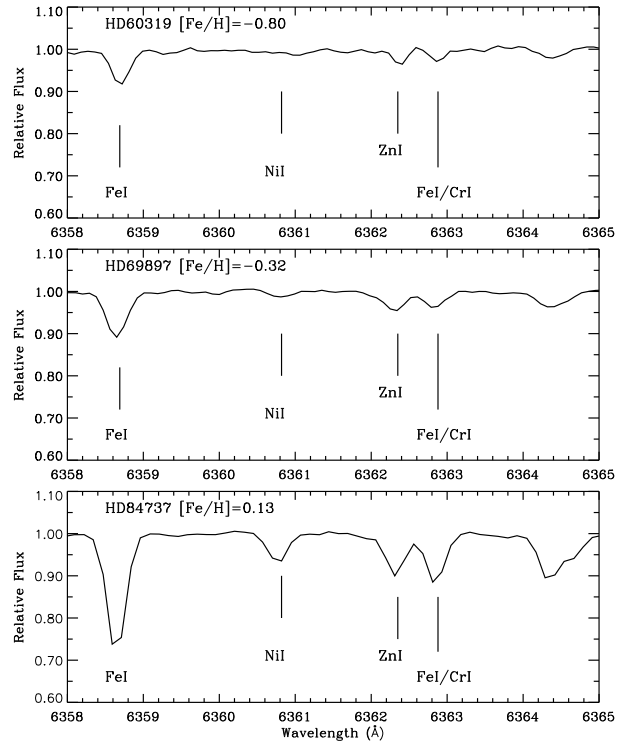}}
\caption{A sequence of spectra around the $\lambda$6362.35\,\AA\ \ZnI\ line
observed with the Coud\'{e} Echelle Spectrograph attached to the
2.16m telescope at the National Astronomical Observatories, Xinglong, China.}
\label{fig:sp.xin}
\end{figure}

\begin{figure}
\resizebox{\hsize}{!}{\includegraphics{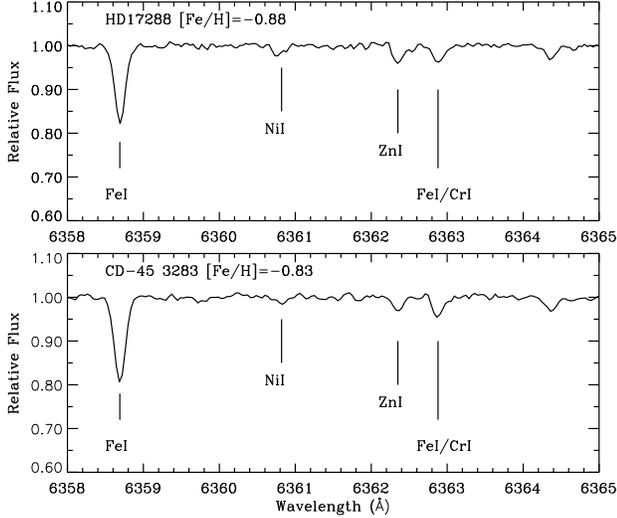}}
\caption{Two spectra around the $\lambda$6362.35\,\AA\ \ZnI\ line
observed with the EMMI spectrograph on the ESO 3.5m telescope, The two
stars have about the same values of \teff\, \logg\ and \feh . Still,
the $\lambda$6362.35\,\AA\ \ZnI\ line appears stronger in \object{HD\,17288}
(a thick disk star) than in \object{CD\,$-45 \, 3283$}
(an alpha-poor halo star).}
\label{fig:sp.eso}
\end{figure}

The second set of stars consists of 19 disk and halo stars with
overlapping metallicities in the range $-1.0 < \feh < -0.4$ taken from
the work of Nissen \& Schuster (\cite{Nissen97}), who determined various
abundance ratios from high resolution
($R$ = 60\,000, $S/N \sim$ 150 - 200) spectra obtained
with the EMMI spectrograph on the ESO 3.5m NTT telescope. The stars were
originally selected from the catalogues of $uvby - \beta$ photometry by
Schuster \& Nissen (\cite{Schuster88}) and Schuster et al. (\cite{Schuster93}),
and have atmospheric parameters in the range $5400 \leq \teff \leq 6300$~K
and $4.0 < \logg < 4.6$. According to the value of the Galactic rotational
velocity component, $V_{rot}$, the stars were classified 
as belonging either to the Galactic halo  ($V_{rot} < 50$\,\kmprs ) or the 
thick disk ($V_{rot} > 150$\,\kmprs ).

For details about the reduction of the spectra we refer to the
papers by Chen et al. (\cite{Chen00}, \cite{Chen02}) and 
Nissen \& Schuster (\cite{Nissen97}). A special problem with the
$\lambda$6362.35\,\AA\ \ZnI\ line is that it lays in the midst of a very broad
and shallow absorption feature identified by Michell \& Mohler
(\cite{Mitchell65}) as due to a \CaI\ auto-ionization line with a
central wavelength of 6361.8\,\AA .
In the solar flux spectrum (Kurucz et al. \cite{Kurucz84}) the central
depth of the line is about 5\% and the total width is more than 15\,\AA .
By fitting a polynomial function to this broad line we have rectified
the spectra around the $\lambda$6362.35\,\AA\ \ZnI\ line. 
Fig.~\ref{fig:sp.xin} shows the resulting spectra for three stars of
different metallicities observed 
with the 2.16m telescope at Xinglong, and Fig.~\ref{fig:sp.eso} shows
two representative spectra observed with the ESO NTT.

The equivalent width of the $\lambda$6362.35\,\AA\ \ZnI\ line was
measured in the rectified spectra by Gaussian fitting
so that the contribution from the nearby FeI/CrI line at 6362.88\,\AA\
can be nearly avoided. The equivalent width for the Sun ($EW = 22.4$\,m\AA ), 
was obtained from a Moon spectrum observed at Xinglong in the same way
as the programme stars. This value is consistent with the 
equivalent width  of the \ZnI\ line obtained from the 
the solar flux spectrum of Kurucz et al. (\cite{Kurucz84}).

\begin{figure}
\resizebox{\hsize}{!}{\includegraphics{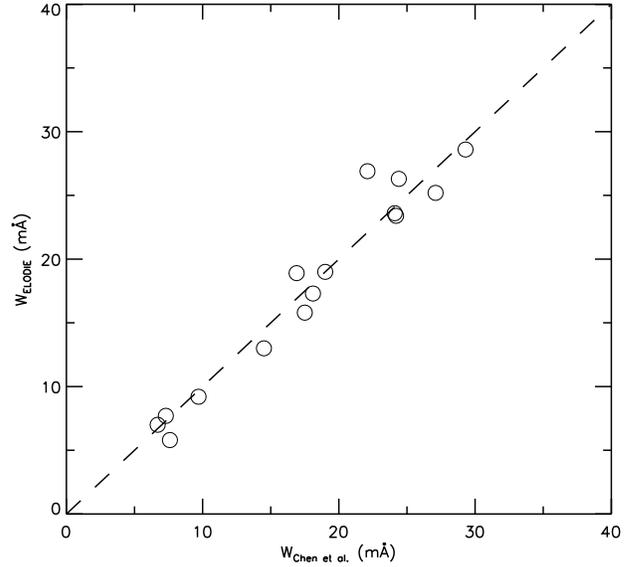}}
\caption{The equivalent width of the $\lambda$6362.35\,\AA\ \ZnI\ line
as measured from ELODIE spectra (Prugniel \& Soubiran \cite{Prugniel01})
versus the equivalent width from our spectra.}
\label{fig:w-w}
\end{figure}

The measured equivalent widths are given in Tables 
\ref{tb:ZnXin} and \ref{tb:ZnEMMI}.
Some of the stars are included in the ELODIE public
library of high resolution spectra (Prugniel \& Soubiran \cite{Prugniel01}).
With a resolution of $R$ = 42\,000 and a  $S/N \sim$ 150 - 200  these
spectra are of similar quality as our spectra and
as a check, equivalent widths of the $\lambda$6362.35\,\AA\ \ZnI\ line
have been measured in the ELODIE spectra also and are listed in 
Tables \ref{tb:ZnXin} and \ref{tb:ZnEMMI}.
Figure \ref{fig:w-w} shows a comparison between the two sets of 
equivalent widths.
As seen the agreement is quite satisfactory. The rms scatter around
the 1:1 line is 1.8\,m\AA\ suggesting that the error of our
equivalent measurements is around 1.3\,m\AA .

\section{Analysis}
\label{analysis}
Stellar parameters, \teff\ and \logg ,
are taken from the above-mentioned papers from which the stars were
selected. In Chen et al. (\cite{Chen00}, \cite{Chen02}),
\teff\ was determined from the $b-y$ colour index using the
IRFM calibration of Alonso et al. (\cite{Alonso96}) and \logg\ was
calculated from the Hipparcos parallax. In Nissen \& Schuster (\cite{Nissen97}),
\teff\ was determined from the excitation balance of \FeI\ lines
but it was checked that the values agree quite well with effective
temperatures derived from $b-y$. The average difference, 
$T_{\rm exc.}-T_{b-y}$, is 65\,K and the rms scatter of the difference
is $\pm55$\,K. As discussed later (see Table~\ref{tb:abuerr}) this 
systematic difference in \teff\ corresponds to a change in \znfe\
of about 0.02\,dex - quite negligible compared to other error sources.
Concerning the gravity parameter, we note that some stars in Nissen \& Schuster 
are too distant to have accurate parallaxes. Hence, \logg\ was determined from
the ionization balance of \FeI\ and \FeII\ lines. Fourteen of the 19 stars
in Table~\ref{tb:ZnEMMI} have, however, Hipparcos parallaxes with a
relative error less than 25\,\%, corresponding to a \logg\ error less than
about 0.20\,dex. The mean difference between the
spectroscopic and parallax-derived gravity is 0.03\,dex with a rms
scatter of $\pm0.15$ dex. The corresponding difference in \znfe\ is 
negligible. Hence, we conlcude that the different ways of deriving 
stellar parameters will 
not introduce significant effects in the abundance determination.
Furthermore, we note that one star that is in common between the
two samples, \object{HD\, 60319} (\,=\,\object{G088-040}), has exactly the 
same abundance ratios in Tables \ref{tb:ZnXin} and \ref{tb:ZnEMMI},
but this is of course a fortuitous agreement.

\begin{table}
\caption{Stellar parameters, equivalent widths of the 6362.35\,\AA\
\ZnI\ line and derived abundance ratios \ZnFe\ and \NiFe\ for
disk stars observed with the Xinglong 2.16m telescope.
$W_a$ is the equivalent width measured from the Xinglong
spectra  and $W_b$ the value obtained from ELODIE spectra.
The stars are classified as belonging to the thin disk 
(D) or the thick disk population (TD) according to their kinematics.}
\label{tb:ZnXin}
\setlength{\tabcolsep}{0.08cm}
\begin{tabular}{rrccrrrrr}
\noalign{\smallskip}
\hline
\noalign{\smallskip}
Star & Pop & \teff & \logg & \feh  & $W_a$ & $W_b$ & \znfe & \nife \\
     &      &   K   &       &       & m\AA   &  m\AA  &       &       \\
\noalign{\smallskip}
\hline
\noalign{\smallskip}
   Sun       &   D  &  5780 &  4.44 &    0.00 &  22.4 &       &     0.00 &     0.00 \\
   \object{HD\,693}  &   D  &  6173 &  4.11 & $-$0.30 &  14.5 &  13.0 &  $-$0.03 &  $-$0.06 \\
   \object{HD\,9826} &   D  &  6119 &  4.12 &    0.12 &  33.7 &       &     0.05 &  $-$0.06 \\
   \object{HD\,10307}&   D  &  5776 &  4.13 & $-$0.12 &  19.7 &       &  $-$0.06 &     0.09 \\
   \object{HD\,10453}&   D  &  6368 &  3.96 & $-$0.46 &  13.0 &       &     0.08 &  $-$0.03 \\
   \object{HD\,13540}&   D  &  6301 &  4.12 & $-$0.43 &  13.9 &       &     0.10 &  $-$0.06 \\
   \object{HD\,16895}&   D  &  6228 &  4.27 &    0.01 &  21.0 &       &  $-$0.09 &  $-$0.05 \\
   \object{HD\,22484}&   D  &  5915 &  4.03 & $-$0.16 &  24.1 &  23.5 &     0.06 &     0.07 \\
   \object{HD\,33632}&   D  &  5962 &  4.30 & $-$0.21 &  17.7 &       &  $-$0.01 &  $-$0.04 \\
   \object{HD\,34411}&   D  &  5773 &  4.02 & $-$0.09 &  29.3 &  28.6 &     0.15 &     0.12 \\
   \object{HD\,39315}&   D  &  6202 &  3.83 & $-$0.38 &  14.5 &       &     0.02 &  $-$0.07 \\
   \object{HD\,39587}&   D  &  5805 &  4.29 & $-$0.18 &  16.9 &  18.9 &  $-$0.08 &  $-$0.04 \\
   \object{HD\,39833}&   D  &  5767 &  4.06 & $-$0.10 &  27.1 &  25.2 &     0.11 &     0.12 \\
   \object{HD\,43947}&   D  &  5859 &  4.23 & $-$0.30 &  16.7 &       &     0.02 &  $-$0.01 \\
   \object{HD\,55575}&   D  &  5802 &  4.36 & $-$0.32 &  15.2 &       &     0.01 &  $-$0.03 \\
   \object{HD\,58551}&   D  &  6149 &  4.22 & $-$0.56 &  14.4 &       &     0.21 &     0.05 \\
   \object{HD\,59380}&   D  &  6280 &  4.27 & $-$0.16 &  21.2 &       &     0.06 &  $-$0.01 \\
   \object{HD\,60319}&  TD  &  5867 &  4.24 & $-$0.80 &   7.9 &       &     0.11 &     0.01 \\
   \object{HD\,68146}&   D  &  6227 &  4.16 & $-$0.15 &  21.6 &       &     0.04 &     0.06 \\
   \object{HD\,69897}&   D  &  6243 &  4.28 & $-$0.32 &  16.4 &       &     0.07 &     0.03 \\
   \object{HD\,75332}&   D  &  6130 &  4.32 &    0.00 &  27.6 &       &     0.03 &  $-$0.09 \\
   \object{HD\,76349}&   D  &  6004 &  4.21 & $-$0.44 &  17.8 &       &     0.18 &  $-$0.01 \\
   \object{HD\,77967}&   D  &  6329 &  4.15 & $-$0.46 &  13.4 &       &     0.10 &  $-$0.03 \\
   \object{HD\,79028}&   D  &  5874 &  4.06 & $-$0.09 &  27.6 &       &     0.09 &     0.08 \\
   \object{HD\,82328}&   D  &  6308 &  3.84 & $-$0.13 &  18.4 &       &  $-$0.06 &  $-$0.12 \\
   \object{HD\,84737}&   D  &  5813 &  4.12 &    0.13 &  28.6 &       &  $-$0.04 &  $-$0.06 \\
   \object{HD\,90839}&   D  &  6051 &  4.36 & $-$0.17 &  15.2 &       &  $-$0.14 &  $-$0.04 \\
   \object{HD\,91889}&  TD  &  6020 &  4.15 & $-$0.25 &  18.9 &       &     0.05 &     0.03 \\
   \object{HD\,94280}&   D  &  6063 &  4.10 & $-$0.06 &  30.6 &       &     0.14 &     0.13 \\
   \object{HD\,95128}&   D  &  5731 &  4.16 & $-$0.07 &  24.2 &  23.5 &     0.01 &  $-$0.02 \\
   \object{HD\,100180}&  D  &  5866 &  4.12 & $-$0.19 &  24.0 &       &     0.11 &     0.05 \\
   \object{HD\,101676}&  D  &  6102 &  4.09 & $-$0.51 &  11.2 &       &     0.02 &     0.03 \\
   \object{HD\,109303}&  D  &  5905 &  4.10 & $-$0.59 &  12.3 &       &     0.12 &  $-$0.07 \\
   \object{HD\,114710}&  D  &  5877 &  4.24 & $-$0.05 &  22.1 &  26.9 &  $-$0.05 &     0.00 \\
   \object{HD\,115383}&  D  &  5866 &  4.03 & $-$0.02 &  24.4 &  26.3 &  $-$0.07 &     0.01 \\
   \object{HD\,150177}&  D  &  6061 &  3.93 & $-$0.63 &   9.7 &   9.2 &     0.03 &  $-$0.02 \\
   \object{HD\,154417}&  D  &  5925 &  4.30 & $-$0.05 &  27.3 &       &     0.08 &  $-$0.04 \\
   \object{HD\,157466}&  D  &  5935 &  4.32 & $-$0.46 &  13.8 &       &     0.08 &     0.06 \\
   \object{HD\,162004}&  D  &  6059 &  4.12 & $-$0.26 &  19.1 &       &     0.04 &     0.07 \\
   \object{HD\,187013}&  D  &  6298 &  4.15 & $-$0.02 &  19.3 &       &  $-$0.12 &  $-$0.10 \\
   \object{HD\,189340}&  D  &  5888 &  4.26 & $-$0.30 &  18.1 &  17.3 &     0.06 &     0.03 \\
   \object{HD\,204363}&  D  &  6141 &  4.18 & $-$0.44 &  12.7 &       &     0.02 &  $-$0.05 \\
   \object{HD\,215648}&  D  &  6158 &  3.96 & $-$0.24 &  17.5 &  15.8 &  $-$0.01 &  $-$0.02 \\
   \object{HD\,216106}&  D  &  5923 &  3.74 & $-$0.15 &  19.5 &       &  $-$0.09 &  $-$0.13 \\
   \object{HD\,222368}&  D  &  6178 &  4.08 & $-$0.13 &  19.0 &  19.0 &  $-$0.05 &  $-$0.03 \\
\hline
\end{tabular}
\end{table}

\begin{table}
\caption{Same as Table \ref{tb:ZnXin} but for stars from Nissen \&
Schuster (\cite{Nissen97}) and $W_a$ is measured 
from the ESO NTT spectra.} 
\label{tb:ZnEMMI}
\setlength{\tabcolsep}{0.08cm}
\begin{tabular}{rrcccrrrr}
\noalign{\smallskip}
\hline
\noalign{\smallskip}
Star & Pop & \teff & \logg & \feh  & $W_a$ & $W_b$ & \znfe & \nife \\
     &      &   K   &       &       & m\AA   &  m\AA  &       &       \\
\noalign{\smallskip}
\hline
\noalign{\smallskip}
    \object{HD\,17288}&   TD  &  5714 &  4.44& $-$0.88 &   7.5 &       &   0.25 &   0.00 \\
    \object{HD\,17820}&   TD  &  5807 &  4.24& $-$0.67 &  11.4 &       &   0.20 &   0.03 \\
 \object{G\,005$-$040}&    H  &  5863 &  4.24& $-$0.83 &   9.4 &       &   0.26 &   0.02 \\
\object{CD\,$-$47\,1087}&   TD  &  5734 &  4.38& $-$0.76 &   9.6 &       &   0.24 &$-$0.01 \\
    \object{HD\,22879}&   TD  &  5851 &  4.36 & $-$0.82 &   7.3 &  7.7  &   0.16 &$-$0.01 \\
    \object{HD\,24339}&   TD  &  5900 &  4.37 & $-$0.63 &  10.3 &       &   0.14 &   0.02 \\
    \object{HD\,25704}&   TD  &  5886 &  4.33 & $-$0.85 &   6.6 &       &   0.12 &$-$0.02 \\
\object{CD\,$-$57\,1633}&    H  &  5933 &  4.26& $-$0.90 &   4.1 &       &$-$0.04 &$-$0.18 \\
\object{CD\,$-$45\,3283}&    H  &  5672 &  4.57& $-$0.83 &   4.1 &       &   0.00 &$-$0.17 \\
    \object{HD\,60319}&   TD  &  5967 &  4.26 & $-$0.80 &   7.1 &       &   0.11 &   0.01 \\
    \object{HD\,76932}&   TD  &  5914 &  4.23 & $-$0.85 &   6.7 &  7.0  &   0.12 &   0.01 \\
 \object{G\,046$-$031}&    H  &  6021 &  4.44 & $-$0.75 &   6.2 &       &   0.04 &$-$0.12 \\
   \object{HD\,103723}&    H  &  6029 &  4.32 & $-$0.79 &   5.2 &       &$-$0.02 &$-$0.11 \\
   \object{HD\,105004}&    H  &  5831 &  4.36 & $-$0.80 &   6.2 &       &   0.06 &$-$0.06 \\
   \object{HD\,106516}&   TD  &  6269 &  4.51 & $-$0.61 &   7.6 &  5.8  &   0.05 &   0.00 \\
   \object{GCRV\,7547}&    D  &  6272 &  4.03 & $-$0.42 &  11.1 &       &   0.01 &   0.00 \\
   \object{HD\,113679}&    H  &  5720 &  4.14 & $-$0.65 &  10.1 &       &   0.18 &   0.03 \\
   \object{HD\,120559}&    D  &  5396 &  4.38 & $-$0.93 &   5.0 &       &   0.12 &$-$0.01 \\
   \object{HD\,121004}&    H  &  5686 &  4.40 & $-$0.70 &   9.4 &       &   0.17 &   0.00 \\
\hline
\end{tabular}
\end{table}

Our determination of abundances is based on 1D model atmospheres
computed with the MARCS code using updated continuous
opacities (Asplund et al. \cite{Asplund97}) and including UV line blanketing
by millions of absorption lines.
LTE is assumed both in constructing the models and in deriving abundances.
For all stars
the microturbulence parameter $\xi_{t}$ was determined by requesting
that the iron  abundance derived from \FeI\ lines should be independent
of the equivalent width. 

For stars selected from Chen et al. (\cite{Chen02}) we have adopted the
\feh\ values given in that paper. These iron abundances are based on
equivalent widths of 18 weak \FeII\ lines analyzed in a differential
way with respect to the Sun. For stars in Chen et al. (\cite{Chen00}),
we redetermined the iron abundances using equivalent widths of a
subset of 8 of the 18 $\FeII$ lines.
The abundances are nearly the same as those presented
in  Chen et al. (\cite{Chen00}) based on \FeI\ lines. For the Nissen \& Schuster
(\cite{Nissen97}) stars we use their \feh\ values as determined
from 104 \FeI\ and 12 \FeII\ lines (forced to give the same iron
abundance due to the way the gravity was determined).

In deriving zinc abundances from the $\lambda$6362.35\,\AA\ $\ZnI$ line
($\chi_{\rm exc} = 5.79$\,eV), we have adopted log$gf = 0.14$
as determined by Bi\'{e}mont \& Godefroid (\cite{Biemont80})
from multi-configurational Hartree-Fock calculations. Using
the MARCS model atmosphere for the Sun and an equivalent width
of 22.4\,m\AA\ of the 6362.35\,\AA\ line as
measured from the Moon spectrum observed in the same way
as the program stars, we derive a solar zinc abundance
of log$\epsilon$(Zn) = 4.52. This is considerably below the
meteoritic Zn abundance of 4.67 $\pm 0.04$
(Grevesse \& Sauval \cite{Grevesse98}). The same problem was
encountered by Bi\'{e}mont \& Godefroid (\cite{Biemont80}),
who derived a solar
photospheric zinc abundance of log$\epsilon$(Zn) = 4.54 from
the $\lambda$6362.35\,\AA\ \ZnI\ line
using the Holweger \& M\"{u}ller (\cite{Holweger74}) model of the Sun.
This difference
between photospheric and meteoritic Zn abundances may well
be due to an  error in the $\log gf$ value of the \ZnI\
line. However, when determining differential Zn abundances with respect
to the Sun, i.e. \znh , using
our solar photospheric abundance of $\log \epsilon$(Zn) = 4.52
as a reference, the possible error in $\log gf$ cancels. The resulting
values of \ZnFe\ are given in Tables \ref{tb:ZnXin} and \ref{tb:ZnEMMI}.
The two tables also include values of \NiFe\ as derived
from more than 20 \NiI\ lines (see Chen et al. \cite{Chen00} and
Nissen \& Schuster \cite{Nissen97})

\section{Errors of the derived abundance ratios}
\label{errors}
The uncertainties of the model parameters are estimated to be 
$\pm 100$\,K in temperature, $\pm 0.15$\,dex in gravity, 
$\pm 0.1$\,dex in metallicity and $\pm 0.3$\,\kmprs\ in
microturbulence. The dependence of the derived \feh\ and \ZnFe\ 
abundance ratios  on the
stellar parameters is calculated by altering temperature, gravity,
metallicity and microturbulence for \object{HD\,60319}, 
\object{HD\,69897} and the Sun,
representing most of our metallicity range. As seen from
Table~\ref{tb:abuerr}, \ZnFe\ is most sensitive to gravity,
but the combined error is small - less than 0.06\,dex if the individual
errors are added quadratically. We also checked that the derived
\ZnFe\ ratio is only marginally affected ($<\!0.02$\,dex), when
alpha-element enhanced models
([$\alpha$/Fe] = +0.4, $\alpha$ = O, Ne, Mg, Si, S, Ca, and Ti)
are used instead of models with [$\alpha$/Fe] = 0.0.
Taking into account the uncertainty of the measured equivalent width
of the $\lambda$6362.35\,\AA\ \ZnI\ line ($\pm$ 1.3\,m\AA ) we then 
estimate a total error of about 0.07\,dex for \ZnFe\ and \feh.
A similar error is estimated for \ZnNi\ and \nih .

\begin{table}
\begin{center}
\caption{Dependence of \feh\ and \ZnFe\ on stellar parameter variations
for two representative stars and the Sun. Zn abundances are
based on the $\lambda6362.35$ \ZnI\ line and Fe abundances on \FeII\ lines}
\label{tb:abuerr}
\setlength{\tabcolsep}{0.05cm}
\begin{tabular}{lrrrrrrrr}
\hline
\noalign{\smallskip}
     & \multicolumn{2}{c}{HD\,60319$^1$}   & \multicolumn{2}{c}{HD\,69897$^2$}
     &\multicolumn{2}{c}{ Sun$^3$}  \\[-1.8mm]
     & \multicolumn{2}{c}{-------------------} & \multicolumn{2}{c}{------------------}
     & \multicolumn{2}{c}{------------------}  \\[-2.0mm]
\noalign{\smallskip}
 & $\Delta \ffe$ & $\Delta \znfe$ & $\Delta \ffe$ & $\Delta \znfe$ &
    $\Delta \ffe$ & $\Delta \znfe$  \\
\noalign{\smallskip}
\hline
\noalign{\smallskip}
$\Delta \teff = 100$~K & $-$0.01 & 0.02 & $-$0.01 &0.03 & $-$0.04 & 0.02 \\
$\Delta \logg= 0.15$ & 0.06 & $-$0.03 & 0.06 & $-$0.03 & 0.06 & $-$0.03 \\
$\Delta \feh$= 0.1 & 0.01 & $-$0.01 & 0.01 & $-$0.01 & 0.03 & $-$0.01 \\
$\Delta \xi_t= 0.3 $ km/s & $-$0.02 & 0.02 & $-$0.04 & 0.03 & $-$0.03 & 0.01 \\
$E_{\gamma} = 1.0 \rightarrow 2.0$ &  & 0.00 &  & 0.00 &  & $-$0.02  \\
\noalign{\smallskip} \hline \noalign{\smallskip}

\multicolumn{7}{l}{1) \object{HD\,60319}: $\teff$=5867~K, 
$\logg$=4.24, $\feh$=$-0.8$, $\xi_t$=1.6}\\
\multicolumn{7}{l}{2) \object{HD\,69897}: $\teff$=6243~K, 
$\logg$=4.28, $\feh$=$-0.3$, $\xi_t$=2.0} \\
\multicolumn{7}{l}{3) The Sun: $\teff$=5870~K, $\logg$=4.44, $\feh$=0.0, $\xi_t$=1.15} \\
\end{tabular}
\end{center}
\end{table}

In calculating zinc abundances from the 6362.35\,\AA\ \ZnI\ line we 
adopted the Uns\"{o}ld (\cite{Unsold55}) approximation to the Van der Waals
interaction constant with an enhancement factor $E_{\gamma} = 1.5$.
As seen from Table \ref{tb:abuerr}, the derived zinc abundance depends
only slightly on the value of $E_{\gamma}$; the effect of increasing
the enhancement factor from 1.0 to 2.0 is less than 0.03\,dex even
in the most metal rich stars. This is in stark contrast to the effect
of $E_{\gamma}$ on the stronger \ZnI\ lines at 4722.2
and 4810.5\,\AA , for which an increase of $E_{\gamma}$
from 1.0 to 2.0 leads to a decrease of the derived Zn abundance for the Sun
with about 0.13\,dex, whereas the effect for the metal-poor disk stars
is a decrease of 0.02\,dex only. Hence, \ZnFe\ in metal-poor disk stars
becomes very sensitive to the adopted value of $E_{\gamma}$ if the stronger
\ZnI\ lines are used. This is the main reason that we have avoided these
lines in the present paper.

As mentioned in Sect. \ref{obs} the  6362.35\,\AA\ \ZnI\ line 
lays in the midst of a $\sim \! 15$\,\AA\ broad and shallow
absorption feature with a central depth of a few percent 
identified by Michell \& and Mohler
(\cite{Mitchell65}) as due to a \CaI\ auto-ionization line.
The equivalent width of the
\ZnI\ line was measured relative to the local apparent continuum
but the small contribution of the \CaI\ auto-ionization line
to the absorption coefficient has been neglected in our analysis.
We have investigated the effect of this auto-ionization line
on the derived Zn abundance by synthesizing the solar flux spectrum
in the spectral region 6356 - 6368\,\AA. In order to fit the large
width, we adopt a very high radiation damping constant 
(1.4\,E+12) of the \CaI\ auto-ionization line and vary its
(unknown) log$gf$ value 
so that the right equivalent width is obtained. In the case of the Sun, 
the effect of including the \CaI\ auto-ionization line
is to increase the Zn abundance by +0.03 dex, i.e. from
log$\epsilon$(Zn) = 4.52 to 4.55.  For  a typical metal-poor thick disk
star, \object{HD\,60319}, with $\feh = -0.80$ and 
$\CaFe \sim +0.2$, the correction of the  Zn abundance is of the order
of +0.01 dex.
Differentially with respect to the Sun, the largest effect of including
the opacity contribution from the \CaI\ auto-ionization line would then be
a decrease of \ZnFe\ for the most metal-poor stars by about 0.02 dex.
This is quite negligible compared with the other error sources.

As described in Sect. \ref{analysis} we have derived the abundances
on the basis of plane parallel, homogeneous (1D) model atmospheres and
the assumption of LTE. Asplund et al. (\cite{Asplund99}) have shown that
there may be significant effects from using three-dimensional (3D)
hydrodynamical models instead, especially for metal-poor stars and
for lines formed in the upper part of the atmosphere such as molecular
lines (Asplund \& Garc\'{\i}a P{\'e}rez \cite{Asplund01}). For lines
formed deep in the atmosphere, such as the \ZnI\ and \FeII\ lines
which we have used, the 3D effects are, however, small 
(Nissen et al. \cite{Nissen04b}) and go in the same direction.
Hence, we don't expect that the derived \ZnFe\ is subject to any significant
3D corrections.

Regarding non-LTE effects on \ZnFe\ we first note that
for the Chen et al. sample the iron abundances are based on \FeII\ lines 
for which departures from LTE are small according to the computations of 
Th\'evenin \& Idiart (\cite{Thevenin99}). 
For the Nissen \& Schuster(\cite{Nissen97}) stars we adopted iron abundances
based on \FeI\ lines and determined the gravities by requiring that
\FeI\ and \FeII\ lines should provide the same Fe abundance. Hence, one might
expect that the derived iron abundances are too low due to 
an overionization of \FeI\ as predicted by Th\'evenin \& Idiart 
(\cite{Thevenin99}) for metal-poor stars. However, as discussed 
in Sect. \ref{analysis}, gravities derived from
Hipparcos parallaxes for a subsample of the  Nissen \& Schuster stars
agree very well with the spectroscopic gravities. Hence, we conclude 
that the derived iron abundances are not significantly different from
the true Fe abundances. This should not be taken as evidence against the
predicted over-ionization of Fe for metal-poor stars. Rather, it
reflects the way the $gf$-values were determined by Nissen \& Schuster,
namely by an ``inverted" abundance analysis of the two bright stars,
\object{HD\,22879} and \object{HD\,76932}, with their parameters and
abundances taken from Edvardsson et al. (\cite{Edvardsson93}).

It is more unclear if there are non-LTE effects on the derived
Zn abundances. With an ionization potential,
$\chi_{\rm ion} (\ZnI)$ = 9.39\,eV, there are approximately equal
numbers of neutral and ionized zinc atoms at the temperatures and
electron pressures of the line forming regions in the atmospheres
of our stars. Hence, an over\-ionization of \ZnI\ relative to LTE,
as is the case for \FeI\ (Th\'evenin \& Idiart \cite{Thevenin99}),
would lead to an underestimate of the abundance of Zn in the LTE
analysis of \ZnI\ lines.
However, such an over-ionization effect on \ZnI\
is likely to be smaller than the equivalent effect
for \FeI ,  since $N_\ZnI \sim N_\ZnII$ whereas $N_\FeI << N_\FeII$,
where N denotes the number density of the atoms and ions.
Furthermore, we note that the 6362.35\,\AA\ \ZnI\ line, on which our
Zn abundances is based, is a weak, high excitation potential
($\chi_{\rm exc} = 5.79$\,eV) line formed deep in the atmosphere,
where departures from LTE tend to be small due to the high
density. Nevertheless, it would be very desirable to perform
a thorough study of non-LTE effects on the determination of Zn 
abundances.

\section{Results and discussion}
\label{results}
Following Fuhrmann (\cite{Fuhrmann98}, \cite{Fuhrmann00}) we 
have used the total space velocity,
$\Vtot = (U^2 + V^2 +W^2)^{\frac{1}{2}}$, with respect to the
Local Standard of Rest (LSR) to classify the stars in
three main populations: i) thin disk stars with $\Vtot < 85$\,\kmprs ,
ii) thick disk stars with $85 < \Vtot < 180$\,\kmprs , and iii)
halo stars with $\Vtot > 180$\,\kmprs . For stars in Table \ref{tb:ZnXin},
the $U,V,W$ velocity components with respect to the LSR are
taken from Chen et al. (\cite{Chen00}, \cite{Chen02}). 
For stars in Table \ref{tb:ZnEMMI} we have updated the velocity components
calculated by Nissen \& Schuster (\cite{Nissen97}) using
Hipparcos (ESA \cite{ESA97}) or Tycho-2 (H{\o}g et al. \cite{Hog00})
proper motions and adopting the most recent values for the
solar motion with respect to the LSR
(($U_{\sun}, V_{\sun}, W_{\sun}$) = ($-10.0$, +5.2, +7.2\,\kmprs )\footnote{
In the present work, $U$ is defined to be positive in the
anticentre direction} (Dehnen \& Binney \cite{Dehnen98}).
The changes with respect to Nissen \& Schuster are relatively
small except for one star, \object{GCRV\,7547} (\object{CD\,$-$39\,\,7674}) that
was classified as a halo star with an unusually high metallicity
($\feh = -0.42$) by Nissen \& Schuster. With the new proper
motion values from the Tycho catalogue, it turns out to be an
ordinary thin disk star with velocity components
($U, V, W$) = (35, 30, 7)\,\kmprs .

\begin{figure}
\resizebox{\hsize}{!}{\includegraphics{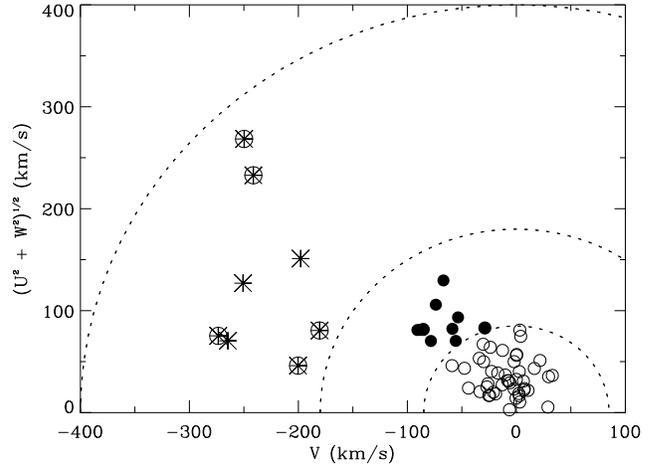}}
\caption{Toomre diagram for stars in Tables \ref{tb:ZnXin} and \ref{tb:ZnEMMI}.
Open circles refer to thin disk stars ($\Vtot < 85$\,\kmprs ),
filled circles to thick disk stars ($85 < \Vtot < 180$\,\kmprs ), 
and asterisks to halo stars ($\Vtot > 180$\,\kmprs ). 
Alpha-poor halo stars are encircled. The dotted circles refer to
total space velocities of 85, 180 and 400\,\kmprs\ with respect to the
Local Standard of Rest.}
\label{fig:toomre}
\end{figure}

The large majority of stars in Table \ref{tb:ZnXin} have
thin disk kinematics, whereas those in Table \ref{tb:ZnEMMI} 
are mainly thick disk or halo stars. As seen from the
Toomre diagram in Fig. \ref{fig:toomre}, the thick disk and the halo
stars are very well separated in $V$ in accordance with the
way the two groups were selected by Nissen \& Schuster (\cite{Nissen97}).
The thin and thick disk stars have a slight overlap in $V$,
but all thick disk stars except one have a rotational
lag in the range $-100 < V < -50$\,\kmprs , whereas all
thin disk stars except one have $V > -50$\,\kmprs .

\begin{figure}
\resizebox{\hsize}{!}{\includegraphics{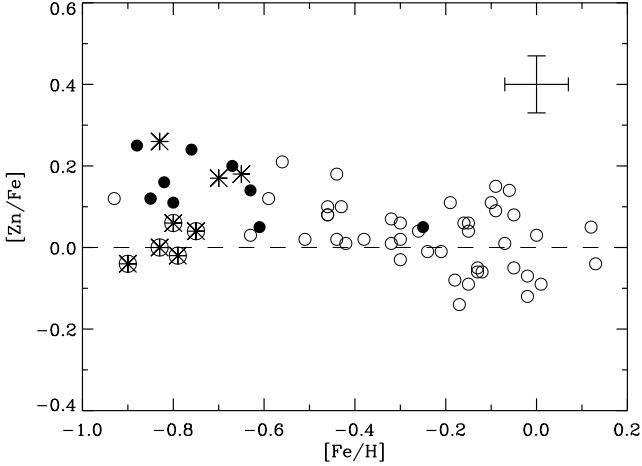}}
\caption{\ZnFe\ versus \feh . The various symbols refer to thin
disk stars (open circles), thick disk stars (filled circles),
and halo stars (asterisks) with alpha-poor halo stars encircled.}
\label{fig:ZnFe}
\end{figure}

Figure \ref{fig:ZnFe} shows \ZnFe\ versus \feh\ with different symbols
for the various populations. Looking first at the thin disk
stars, we see a slight increasing trend of \ZnFe\ with decreasing
metallicity; a straight line least squares fit to the disk star data
gives:
\begin{eqnarray}
\ZnFe = -0.01 \,(\pm 0.02) - 0.16 \,(\pm 0.05) \cdot \feh \nonumber
\end{eqnarray}
with a reduced chi-square, $\chi^2_{red}$ = 1.11, when the estimated
error of 0.07\,dex on \ZnFe\ and \feh\ is adopted. This relation
corresponds to $\ZnFe \simeq +0.1$ for a thin disk metallicity of
$\feh = -0.6$, and the relation agrees quite well with the trend
of \ZnFe\ versus \feh\ found by Reddy et al. (\cite{Reddy03}) 
and Bensby et al. (\cite{Bensby03}) for thin disk stars.

\begin{figure}
\resizebox{\hsize}{!}{\includegraphics{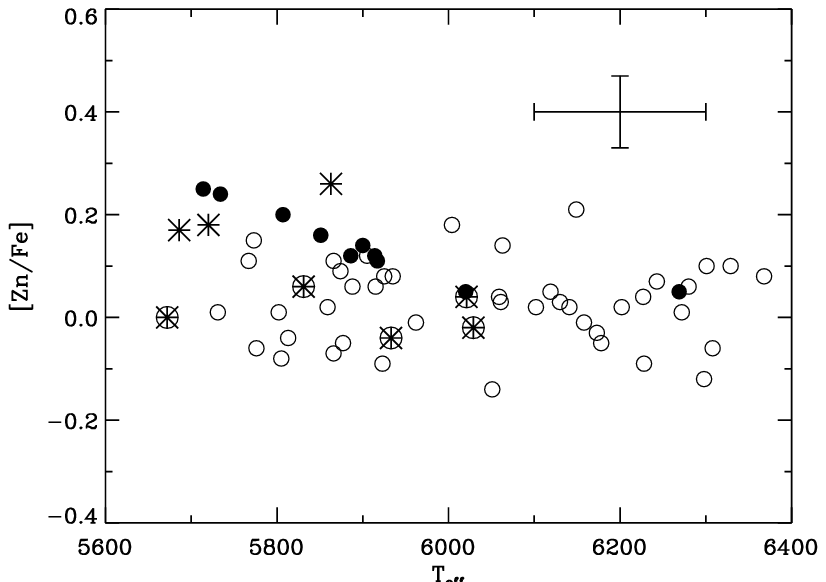}}
\caption{\ZnFe\ versus \teff . The symbols are the same as those
in Fig.~\ref{fig:ZnFe}.}
\label{fig:ZnFeTe}
\end{figure}

\begin{figure}
\resizebox{\hsize}{!}{\includegraphics{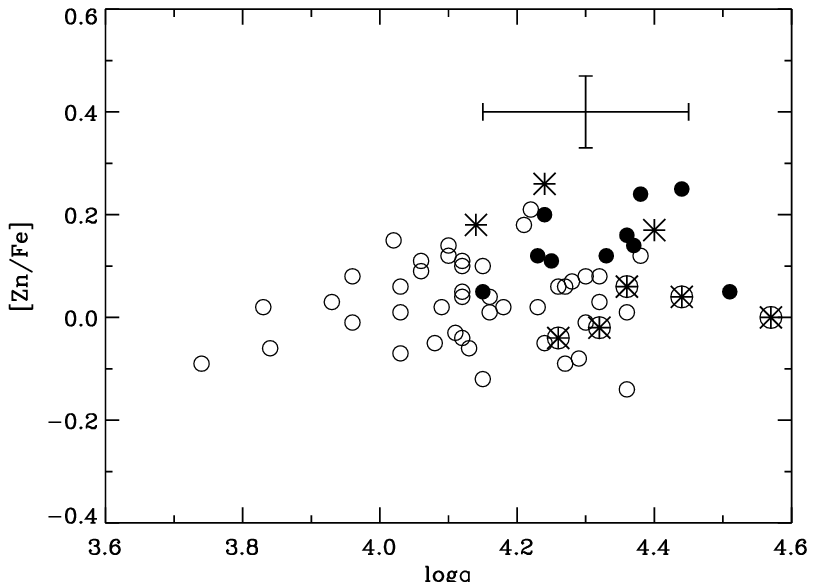}}
\caption{\ZnFe\ versus \logg . The symbols are the same as those
in Fig.~\ref{fig:ZnFe}.}
\label{fig:ZnFeLg}
\end{figure}

Bensby et al. (\cite{Bensby03}) have found evidence of a 
difference of about 0.1\,dex in \ZnFe\ between thick and thin disk 
stars in the metallicity range $-0.6 < \feh < -0.3$. As seen from
Fig. \ref{fig:ZnFe}, we have no thick disk stars in this
metallicity range. Our ten thick disk stars
in the metallicity range  $-0.9 < \feh < -0.6$ have a mean
\ZnFe\ of 0.15\,dex, which is only slightly higher than the value
predicted from the thin disk relation given above.
The explanation may be that our thick disk stars lie in a 
metallicity range, $\feh < -0.6$, where the thick and the thin
disk are not chemically well separated. The same is seen for e.g. the ratio
between alpha-capture elements and Fe, i.e.
[$\alpha$/Fe]. In the metallicity range $-0.6 < \feh < -0.3$
[$\alpha$/Fe] is different for thick and thin disk stars
(Bensby et al. \cite{Bensby03}, \cite{Bensby04}), whereas
the two populations  start to merge together in [$\alpha$/Fe]
for $\feh \simeq -0.7$.

Among the eight halo stars shown in Fig. \ref{fig:ZnFe},
three have an enhanced Zn/Fe ratio
at the same level as the thick disk stars, but the other five
have a solar Zn/Fe ratio. As shown in Figs.~\ref{fig:ZnFeTe} and
~\ref{fig:ZnFeLg}, the eight halo stars and the ten thick disk
stars have similar temperatures and 
gravities, and there is no significant trend
of \ZnFe\ with the stellar parameters. Thus, the scatter of \ZnFe\
among the halo stars seems to be real. Furthermore, the five
halo stars with low \ZnFe\  are among 
the group of so-called alpha-poor halo stars that 
were found by Nissen \& Schuster (\cite{Nissen97})
to have a solar-like ratio between the
alpha-capture elements (O, Mg, Si, Ca and Ti) and Fe.
Hence, Zn tends to follow the  alpha-capture elements
in two respects: by being overabundant in metal-poor disk stars
and by having a solar-like \ZnFe\
for the alpha-poor halo stars. The amplitude of the
\ZnFe\ variations is, however, smaller than in the case of
\OFe\ and \MgFe , i.e. $\sim \! 0.15$\,dex instead of 
$\sim \! 0.3$\,dex.

\begin{figure}
\resizebox{\hsize}{!}{\includegraphics{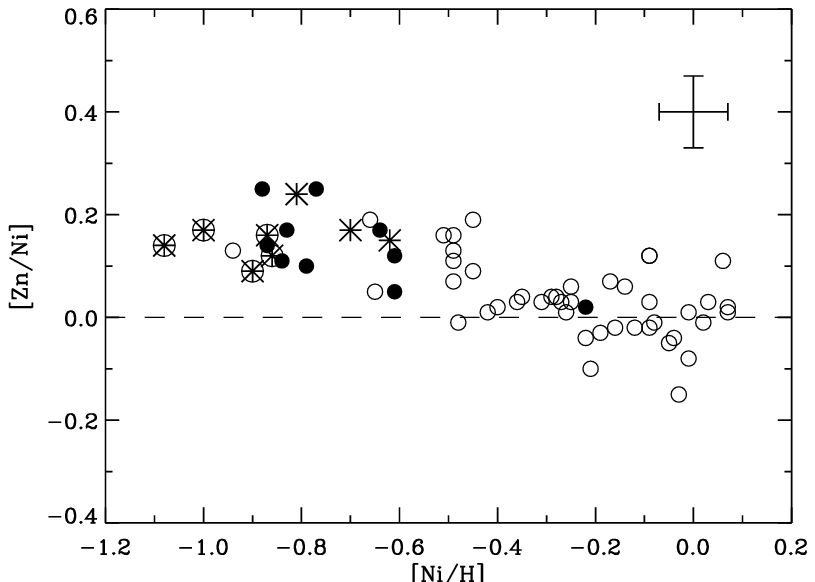}}
\caption{\ZnNi\ versus \nih\ with the same symbols as in Fig.\,\ref{fig:ZnFe}.}
\label{fig:ZnNi}
\end{figure}

Nissen \& Schuster (\cite{Nissen97}) also found
the alpha-poor halo stars to have unusual low abundances
of Na and Ni with respect to Fe. \NaFe\ varies from $-0.4$ to 0.0
and \NiFe\ varies from $-0.2$ to 0.0 with a tight
correlation between the two ratios. As discussed by
Nissen (\cite{Nissen04a}) recent abundance studies
of giant stars in dwarf spheroidal galaxies
(Shetrone et al. \cite{Shetrone03}) show that these stars also
have underabundant values of \NaFe\ and \NiFe\ and
fit the \NaFe\ - \NiFe\ relation of alpha-poor halo stars
very well. This supports the suggestion of Nissen \& Schuster 
(\cite{Nissen97}) that the alpha-poor halo stars have been
accreted from dwarf galaxies with a different star formation
history and/or initial mass function than the Milky Way.
The deficiency of Na and Ni may be
connected to the fact that the yields of both Na and the dominant
Ni isotope ($^{58}$Ni) depend upon the neutron
excess (Thielemann et al. \cite{Thielemann90}). The deficiency of Zn
in the alpha-poor halo stars may then be explained by the
fact that the production of Zn in Type II SNe also depends
on the neutron excess (Timmes et al. \cite{Timmes95}).
Indeed, Fig. \ref{fig:ZnNi} shows that the alpha-poor halo stars
do not stand out in a plot of \ZnNi\ versus \nih .

\begin{figure*}
\centering
\includegraphics[width=17cm]{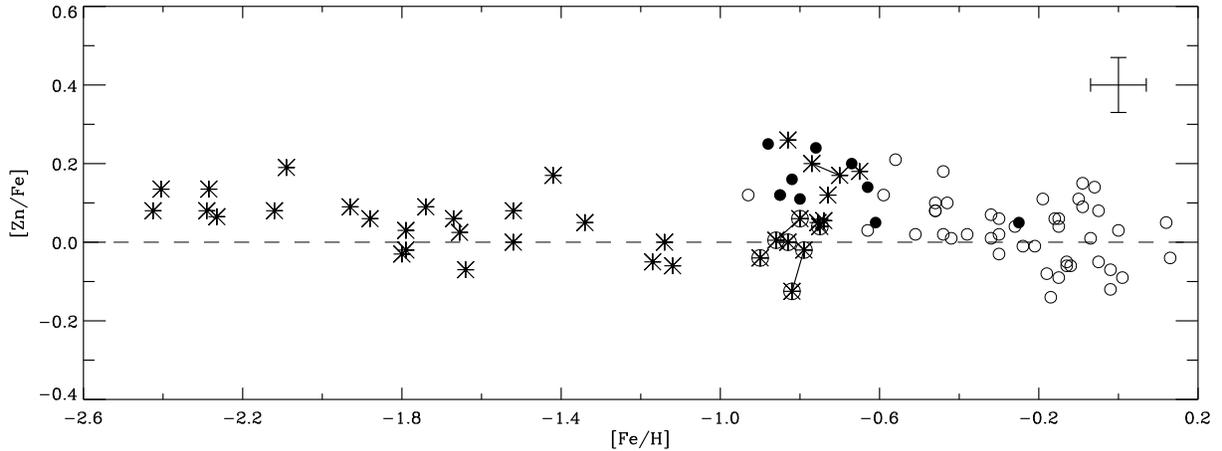}
\caption{\ZnFe\ versus \feh\ with the same symbols as in Fig.\,\ref{fig:ZnFe}
and including \ZnFe\ for metal-poor halo stars as derived from the 
$\lambda \lambda$\,4722.16, 4810.54\,\AA\ \ZnI\ lines by
Nissen et al. (\cite{Nissen04b}). Three stars that are in common between
the present work and Nissen et al. have been connected by straight lines.}
\label{fig:ZnFe.all}
\end{figure*}

The $\lambda$\,6362.25\,\AA\ \ZnI\ line is too weak to be
used for deriving Zn abundances in stars with $\feh < -1.0$. For
such stars the $\lambda \lambda$\,4722.16, 4810.54\,\AA\ \ZnI\ lines
are, however, quite ideal having equivalent widths ranging
from a few m\AA\ to about 50\,m\AA\ when \feh\ increases from
$-2.5$ to $-0.8$. Figure \ref{fig:ZnFe.all} includes such
\ZnFe\ data for 29 stars from Nissen et al. (\cite{Nissen04b}).
Three of the stars, \object{HD\,103723}, \object{HD\,105004}
and \object{HD\,121004}
are in common with the present paper; their symbols are
connected with straight lines in Fig.\,\ref{fig:ZnFe.all}.
As seen, there is a reasonable good agreement between the
two sets of data. The mean difference in \ZnFe\
(Nissen et al. $-$ present paper) is $-0.06$\,dex
suggesting perhaps a slight offset of the data based
on the $\lambda \lambda$\,4722.16, 4810.54\,\AA\ lines,
which may occur because the lines are quite strong in
the Sun causing \ZnFe\ to be sensitive to the
adopted value of the damping constant as discussed in 
Sect.\,\ref{errors}.

From Fig.\,\ref{fig:ZnFe.all} it is seen that although 
the overall \ZnFe\ trend is quite flat over the metallicity range
$-2.5 < \feh < +0.2$, subtle deviations of \ZnFe\ from zero 
seem to occur. 
The most metal-poor halo stars have $\ZnFe \sim +0.1$ and 
metal-poor thin disk and thick disk stars have $\ZnFe \sim$
+0.10 to +0.15. One might argue that such
small deviations could be due to errors in the
analysis, because non-LTE effects were not taken into account. 
On the other hand, it seems very difficult to explain the
systematic difference of about 0.15\,dex in \ZnFe\ between 
thick disk stars and alpha-poor halo stars at the
{\em same metallicity} as a non-LTE effect.
We are here comparing abundances
of stars with nearly the same atmospheric parameters,
\teff , \logg\ and \feh . Hence, any non-LTE effects are
expected to cancel when determining the difference in
\ZnFe . 

We therefore think that the small deviations of \ZnFe\ from
zero are real and contain important information about the
complicated nucleosynthesis of Zn. For the most metal-poor
halo stars Cayrel et al. (\cite{Cayrel04}) also find a slight
overabundance of \ZnFe\ at $\feh = -2.5$ increasing to 
$\ZnFe \sim +0.5$ at $\feh = -4.0$. As discussed by 
Umeda \& Nomoto (\cite{Umeda02}) this trend and other 
non-solar abundance ratios for
the iron-peak elements may be explained with high-energy
(hypernovae) models for core collapse explosions of massive
Population III stars. In the case of the thick disk and metal-poor
thin disk stars, the overabundance of Zn/Fe indicates that there
is a source of zinc production in addition to the weak
$s$-process and alpha-rich freezeout in Type II SNe.
This additional contribution may 
be due to zinc production in the neutrino-powered wind of 10 
to 20 solar mass SNe as suggested by Hoffman et al. (\cite{Hoffman96}),
essentially at the same site as proposed for the $r$-process
(Woosley et al. \cite{Woosley94}).
In this connection, we note that the $r$-process elements are also known 
to be overabundant with respect to Fe in thick disk and metal-poor
thin disk stars (Mashonkina et al. \cite{Mashonkina03}).
Furthermore, it is interesting that it is the $^{64}$Zn isotope,
which is produced in the neutrino-powered wind according to
Hoffman et al. (\cite{Hoffman96}). $^{64}$Zn is the
dominant isotope in the solar system (Anders \& Grevesse \cite{Anders89})
but is underproduced by a factor of three in traditional
calculations of nucleosynthesis of Zn in type II SNe 
(Timmes et al. \cite{Timmes95}, Fig. 3).

As seen from Fig.\,\ref{fig:ZnFe.all}, the halo stars in the
metallicity range $-1.8 < \feh < -1.0$ have $\ZnFe \sim 0.0$.
They form a sequence in the figure that fit well to the
alpha-poor halo stars at $\feh \sim -0.8$. The thick disk
and the halo stars at $\feh \sim -0.8$ with 
$\ZnFe \sim +0.15$ are apparently subject to a different
evolution in \ZnFe . The explanation may be that the Galactic
halo consists of two components as also suggested by
Gratton et al. (\cite{Gratton03}): A dissipative component
with high \ZnFe\ and [$\alpha$/Fe] connected to the thick disk
and an accreted component with low \ZnFe\ and [$\alpha$/Fe].

Although these possible variations of \ZnFe\ as a function of \feh\
are of importance when interpreting abundance data for 
DLAs, we note that the amplitude of \ZnFe\ is smaller
than in the case of [$\alpha$/Fe]. Sulphur is of particular
interest in this connection, because S like Zn is practically
undepleted onto dust so that the observed S/Zn interstellar 
gas ratio in DLAs equals the intrinsic abundance ratio between 
the two elements. As shown by Nissen et al. (\cite{Nissen04b}),
the overabundance of S in halo stars corresponds to
$\SFe \sim +0.3$ to +0.4\,dex. Hence, when plotting \SZn\
versus \znh\ we still get a trend that looks like
\SFe\ versus \feh\ (see Nissen et al. \cite{Nissen04b}, Figs. 
6 and 8). This suggests that \SZn\ may be used
as a ``chemical clock'' to date the star formation process 
at high $z$ albeit with a lower sensitivity than \SFe . 
As discussed by Fenner et al. (\cite{Fenner04}), a better
understanding of the nucleosynthesis of Zn is, however, needed
in order to derive the past history of star formation in DLAs
from the observed S/Zn ratio.

\section{Conclusions}
\label{conclusions}
We have determined Zn abundances for 62 dwarf stars with 
metallicities in the range $-1.0 < \feh < +0.2$. The abundances
are based on equivalent widths of the weak $\lambda 6362.35$\,\AA\
\ZnI\ line and are relatively insensitive to possible errors
in the atmospheric parameters and the damping constant of the
line; thus they are more reliable than Zn abundances based on the
stronger $\lambda \lambda$\,4722.16, 4810.54\,\AA\ \ZnI\ lines,
which have been applied in other studies of Zn abundances.

The stars were grouped into thin disk, thick disk and halo populations 
according to their kinematics. It is found that \ZnFe\ in thin disk
stars shows a slight increasing trend with decreasing metallicity
reaching a value of $\ZnFe \sim \! +0.1$ at $\feh = -0.6$ in agreement
with Reddy et al. (\cite{Reddy03}) and Bensby et al. (\cite{Bensby03}).
Ten thick disk stars in the metallicity range $-0.9 < \feh < -0.6$ have an
average $\ZnFe\ = +0.15$\,dex, whereas five alpha-poor halo
stars in the same metallicity range have $\ZnFe\ \simeq 0.0$\,dex.
Interestingly, the same five stars are also underabundant in Ni
having an average $\NiFe \simeq -0.13$. 

These results suggest that Zn is not an exact tracer of Fe as often
assumed in abundance studies of DLA systems. The overabundance of Zn/Fe 
in metal-poor thin and thick disk stars may be explained by
Zn production in the neutrino-powered wind of 10
to 20 solar mass SNe as suggested by Hoffman et al. (\cite{Hoffman96}).
The deviations of \ZnFe\ from zero are, however, considerably smaller
than in the case of \OFe , \MgFe\ and \SFe ; 
\SZn\ versus \znh\ for DLA systems may therefore still be used to obtain
information on the star formation history in these objects if the
nucleosynthesis of zinc can be better understood.
 
In order to improve on the study of the \ZnFe\ - \feh\ trend for
Galactic stars even more accurate data for the equivalent width of
Zn lines should be obtained in larger samples of disk and halo stars.
At the same time a thorough study of non-LTE effects on the 
determination of Zn abundances is needed, as well as 
further studies of the nucleosynthesis of zinc in various types
of objects.

\begin{acknowledgements}
This work is partly supported by the National Natural Science Foundation
of China and grant NKBRSF G1999075406, and by the Danish Natural Science 
Research Council, grant 21-01-0523.
\end{acknowledgements}

{}
\end{document}